\documentclass[a4paper,fleqn,usenatbib]{mnras}

\usepackage{newtxtext,newtxmath}
\usepackage[T1]{fontenc}
\usepackage{ae,aecompl}
\usepackage{pdflscape}

\usepackage{graphicx}	% Including figure files
\usepackage{amsmath}	% Advanced maths commands
\newcommand{\oversim}[2]{\protect{\mbox{\lower0.5ex\vbox{%
  \baselineskip=0pt\lineskip=0.2ex
  \ialign{$\mathsurround=0pt #1\hfil##\hfil$\crcr#2\crcr\sim\crcr}}}}}
\title[New radial velocities for 40 nearby dwarf galaxies]
{New radial velocities for 40 nearby dwarf galaxies}

\author[Karachentsev, Chazov \& Kaisin]
{Igor D. Karachentsev$^{1}$\thanks{idkarach@gmail.com},
  Maxim I.~Chazov$^{1}$ and Serafim. S.~Kaisin$^{1}$ 
\\
$^1$Special Astrophysical Observatory, The Russian Academy of Sciences, Nizhnij Arkhyz, Karachai-Cherkessian Republic
369167, Russia\\
}

\pubyear{20xx}
\begin{document}
\label{firstpage}

\maketitle

\begin{abstract}
  The 6-meter BTA telescope has been used to determine radial velocities for 40 galaxies,
recently identified in the DESI Legacy Imaging Surveys as nearby objects. Half of them
have kinematic distances within 11 Mpc being new probable companions to the bright 
Local Volume galaxies:  NGC\,628, Maffei\,2, NGC\,2787, M\,81, NGC\,4605 and NGC\,4631.
Six relatively isolated objects with heliocentric velocities  in the range of $[-150, +70]$~km~s$^{-1}$,
together with the blue compact dwarf NGC 6789,
form a diffuse association of dwarf galaxies located in the near part of the Local Void.
\end{abstract}


\section{Introduction} The radial velocity is one of the most important characteristics of galaxies. In the last 
decades, there has been a rapid increase in the number of galaxies with measured radial velocities. Thanks to massive spectral 
surveys of the sky in the optical and radio bands (SDSS, \citep{Aba2009}; ALFALFA, \citep{Hay2011}; HIPASS, \citep{Kor2004}; FASHI, \citep{Zha2024}), 
the number of galaxies with certain velocities has exceeded one million. Nevertheless, many nearby dwarf galaxies, 
especially of low surface brightness, remain with unknown radial velocities. This is particularly the case 
galaxies at high declinations (Dec. $>+38^{\circ}$), not covered by wide-field sky surveys in the neutral hydrogen 21~cm line.

 Many research groups have been searching for and studying nearby galaxies \citep{Chi2013,Mul2018,Sme2018,Mul2019,Crn2019,
Oka2019,Ben2020,Dav2021,Tru2021,Mut2022,Car2022,Mcq2024,Mut2024,Cro2024}.
At present, there are about 1500 galaxies in the Local Volume (LV) of the Universe with a radius of 11 Mpc, whose data have been collected in the database 
Updated Nearby Galaxy Catalog \citep{Kar2013}, an updated version of which is available at 
http://www.sao.ru/lv/lvgdb. Radial velocities have so far been measured for less than 60\% of this sample. Increasing data 
on the velocities of nearby galaxies will allow us to make a clearer picture of virial motions and cosmic flows in the LV, which are determined by the dark matter topography.
 
 Using data from DESI Legacy Imaging Surveys \citep{Dey2019}, we have identified about a hundred galaxy candidates in the LV 
located both in known nearby groups and in the common field between groups \citep{Kar2022a,Kar2024}. Most of them are late-type 
dwarf galaxies (Irr, Im, BCD) with active star formation. The measurement of radial velocities of these galaxies is the goal of our present work.
 
  \begin{figure*}[h]
     \centering
     \includegraphics[width=\textwidth]{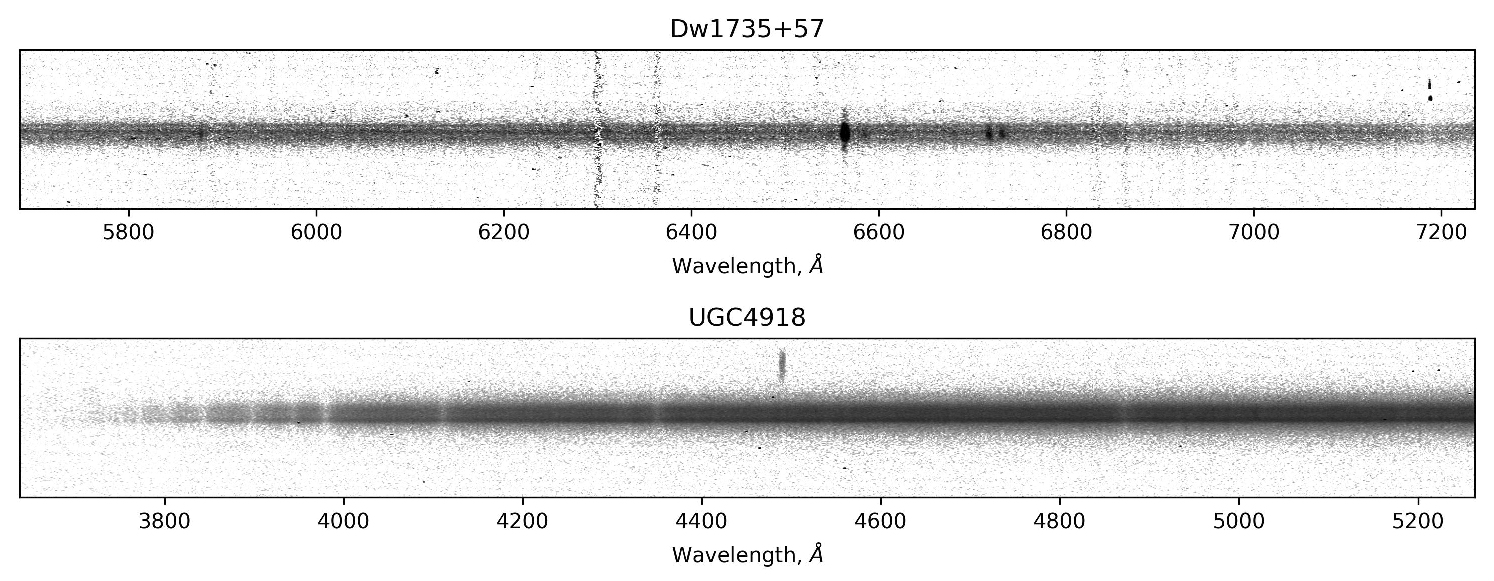}
     \caption{Reproduction of the spectrum of a galaxy with emission lines (Dw1735+57) and a galaxy with absorption lines (UGC\,4918).}
 \end{figure*}

 \section{Observations with the 6-meter BTA telescope}
 Spectral observations at the 6-m telescope were performed with the SCORPIO-1 and SCORPIO-2 devices \citep{Afa2011} in the 
long slit version. The gratings mainly used in the observations for SCORPIO-1 were VPHG1200B and VPHG1200R, providing spectral 
resolution $\Delta\lambda\sim6$\AA\, in the range 3600--5400\AA\, and 5700--7500\AA. Observations for SCORPIO-2 was mainly done 
using VPHG1200@540 grating  in the spectral range 3650-7300\AA. The slit width and length corresponded to $(1.0-1.2)^{\prime\prime}\times 5^{\prime}$. 
The observations were made in the years 2021--2024 at images of (1--2)$^{\prime\prime}$ with exposures of (600--2400)~sec.

The data reduction was done using the standard technique implemented in an IDL-based program, described in detail in previous 
papers (for example, \citet{Ego2018}. For each of the objects studied, we obtained one-dimensional spectra by summing the signal 
in regions corresponding to the sizes of the galaxies. We used the Gaussian approximation to measure the redshifts of the lines, 
correcting the results by applying linearization shifts measured by the nearest emission lines in the night sky. A typical accuracy 
of the radial velocity measurements was $\sim10$~km~s$^{-1}$ for emission galaxies and $\sim45$~km~s$^{-1}$ for objects with absorption lines.

 \section{Results} 
The reproduction of spectra of the galaxy Dw1735+57 with emission lines and the galaxy UGC\,4918 with absorption-only lines are 
shown in Fig. 1. The results of the measured radial velocities for 40 galaxies are presented in Table 1. The columns of the Table 
contain: (1) the name of the galaxy as it appears in the database http://www.sao.ru/lv/lvgdb; (2) the equatorial coordinates at 
epoch J(2000.0); (3) the apparent $B$-magnitude of the galaxy, determined from $g$ and $r$-magnitudes from DESI survey  
as $B=g+0.313(g-r)+0.227$ according to Lupton \footnote{http://www.sdss3.org/dr10/algorithms/}; for some galaxies of low surface 
brightness with patchy structure, the $B$-magnitude estimates are made visually and have an error of $\sim0.5$~mag; (4) the heliocentric 
velocity of the galaxy in km~s$^{-1}$; (5) the radial velocity (km~s$^{-1}$) relative to the centroid of the Local Group; (6) the presence 
of emission or absorption-only lines in the spectrum; (7) the kinematic distance of the galaxy in Mpc, determined in the Numerical Action 
Method (NAM) model taking into account the local flow towards the Virgo cluster and the Local Void expansion \citep{Sha2017,Kou2020}.

 In determining the kinematic distances of the galaxies given in column (7), we made a correction of $\Delta D=+0.3$~Mpc. This value 
was obtained from a comparison of $D_{\rm NAM}$ distances with accurate TRGB-distances for 430 LV galaxies estimated via the tip of 
the red giant branch \citep{Kar2025}. As noted in that paper, the typical error in determination of NAM distances in the LV is 1.6~Mpc 
if the galaxies are located outside the $\sim25^{\circ}$ radius zone around the Virgo cluster. All of the galaxies observed, except for 
two galaxies (Dw1238+33 and SMDG1241+35) fulfil this condition.
 
 \section{Discussion} Among the 40 observed dwarf galaxies, 16 objects have radial velocities in the range of 
$V_{\rm LG} =1000-3700$~km~s$^{-1}$. These galaxies, mostly with emission spectra, are located in the general field outside the Local 
Volume. Some of them (Dw0742+77 and Dw0831+77, VV507 and Dw1707+51) are associated with each other, having similar radial velocities. 
Of the remaining 24 dwarf galaxies, most are likely members of the known local groups. 

 \subsection{New members of the Local Volume groups}
 
 {\em Dw0136+1628, NGC\,628dwA}. \citet{Car2022} determined distances of these galaxies from the surface brightness fluctuations (sbf) 
as 9.45~Mpc and 10.12~Mpc, respectively, which are consistent with the distance of the spiral galaxy NGC\,628, 10.19~Mpc, determined from 
the tip of the red giant branch (TRGB) by \citet{Jan2014}. The radial velocities of both the dwarfs are also quite typical of satellites 
of the host galaxy NGC\,628, which has $V_{\rm LG} = 827$~km~s$^{-1}$.
 
 {\em KKH\,11}. It is a probable member of the Maffei group in a zone of strong Galactic extinction. Our estimate of its radial velocity 
$V_h=111$~km~s$^{-1}$ on the hydrogen and oxygen lines is very different from the value $V_h=269\pm 2$~km~s$^{-1}$ derived from the 21~cm 
line \citep{Beg2008}. \citet{Sch2019} give for  KKH\,11 a value of $V_h=76\pm 2$~km~s$^{-1}$. A reason for these differences may be due to 
the superposition of Galactic hydrogen emission.
 
 {\em PGC\,025409, UGC\,4918}. These are blue compact dwarfs (BCD), possible peripheral members of an association around the spiral galaxy 
NGC\,2787, which has a radial velocity of $V_{\rm LG} = 842$~km~s$^{-1}$.
 
 {\em Dw0921+75.} A probable companion of the dwarf galaxy UGC\,4945 = KDG\,54, which has a radial velocity of $V_{\rm LG} = 833$~km~s$^{-1}$.
 
 {\em SMDG0956+82, UGC\,6451.} New peripheral members of the M81 group. The granular structure of both the dwarfs is consistent to their 
distance of $D\sim 3-4$~Mpc.
 
 {\em KDG\,74, Dw1234+76.} These dSph galaxies with granular structure are possibly located at the far boundary of the M\,81 group. They are 
not detected in the CALEX UV survey \citep{Gil2007}. The radial velocities of both dwarf galaxies are slightly higher than the average velocity 
of galaxies in this group. To confirm the membership of KDG\,74 and Dw1234+76 in the M\,81 group, it is necessary to 
measure their distances.
 
 {\em [KK\,98]121.} It is a dSph satellite of the galaxy NGC\,4051, which has a radial velocity of $V_h=700$~km~s$^{-1}$ and a distance of 
11.0~Mpc by the Tully-Fisher relation \citep{Tul2013}.
 
 {\em KDG\,162.} A dwarf galaxy of transitional type (Tr) between dIrr and dSph with a granular structure. It is a possible companion to the 
peculiar Sm galaxy NGC\,4605 ($V_h=151$~km~s$^{-1}$, $D_{\rm TRGB}=5.55$~Mpc) or a foreground object with $D\simeq 3-4$~Mpc.
 
 {\em SMDG1241+35.} A probable companion to the spiral galaxy NGC\,4631 ($V_h=583$~km~s$^{-1}$, $D_{\rm TRGB}=7.35$~Mpc) or an isolated dwarf 
in the general field.
 
 {\em Dw1245+61.} A transitional type dwarf with a granular structure, not detected in the GALEX survey.  A probable satellite of the galaxy 
NGC\,4605. Its radial velocity from absorption lines with a low signal-to-noise ratio is  measured uncertainly.
 
 \subsection{New dwarf galaxies in the Local Void.} The local cosmic void discovered by \citet{Tul1988} extends from the boundary of the Local 
Group to a distance of $\sim20$~Mpc towards the northern supergalactic pole (RA$\sim 18^h55^m$, Dec$\sim16^{\circ}$). This region is characterised 
also by significant light extinction, making it difficult to find even the nearest galaxies. So far, five nearby galaxies have been known: 
NGC\,6503, KK\,242, UGC\,11411, NGC\,6789 and KK\,246, which are located at high supergalactic latitudes SGB$>+30^{\circ}$. \citet{Don2005} 
and \citet{Sch2019} found two more radio sources in this polar region, HIZOAJ1914+10 and EZOAJ2129+52, with low radial velocities, which are 
not identified with any optical counterparts because of very strong extinction. The distances to these objects, estimated from the radial 
velocity  in the Numerical Action Method 
(NAM) or by the baryonic Tully--Fisher (TF) relation, remain highly uncertain. Our present observations add six more nearby dwarf 
galaxies to the Local Void population: Dw1558+67, Dw1559+46, Dw1645+46, Dw1735+57, Dw1907+63  and Dw1940+64.
 \begin{figure}
    \centering
    \includegraphics[width=0.5\textwidth]{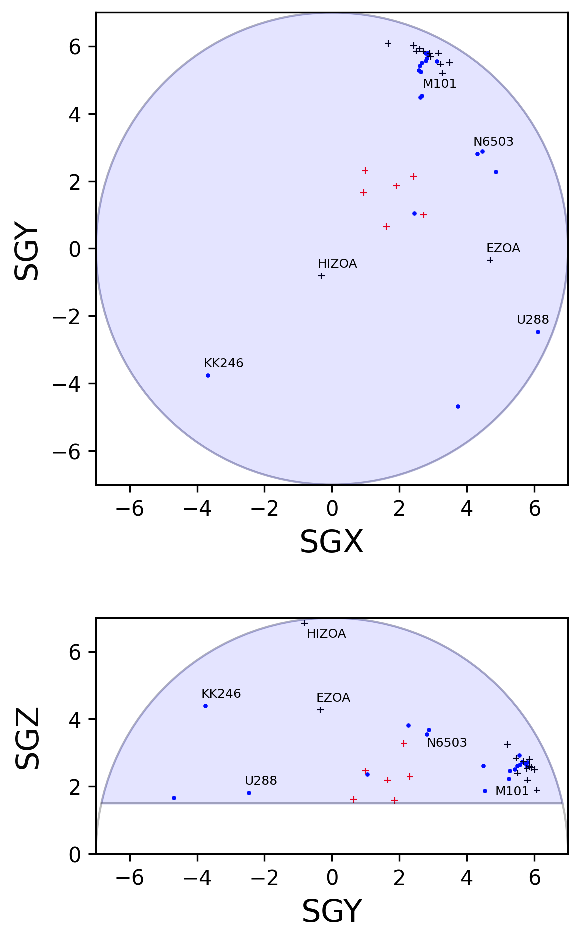}
     \caption{Distributions of 38 galaxies from Table 2  given in two projections: $\{$SGX, SGY$\}$ and $\{$SGY, SGZ$\}$. Galaxies 
with accurate distance estimates (trgb, sbf) are shown with blue circles, while objects with distances of low accuracy (NAM, TF, mem) 
are marked with  black crosses. The new dwarf galaxies detected by us are highlighted in red.}
  \end{figure}

Table 2 represents the 38 known galaxies located in the near part of the Local Void with distances $D<7.0$~Mpc that are elevated above 
the Local Sheet plane (SGZ$ >1.5$ Mpc). Among them, 22 galaxies belong to members of a group around the giant spiral galaxy M\,101 at the 
far edge of the considered volume. The columns of Table 2 contain: (1) object name; (2,3) equatorial coordinates in degrees; (4,5) 
supergalactic coordinates in degrees: (6) the distance in Mpc; (7) the method by which the distance is estimated (``mem‘’~--- by galaxy
 membership in the group); (8--10) supergalactic coordinates in Mpc; (11) -- references to the distance estimates. The six new members of the 
Local Void are placed at the end of Table 2. Distributions of these 38 galaxies in two projections $\{$SGX, SGY$\}$ and $\{$SGY, SGZ$\}$ 
are shown in Fig. 2. Galaxies with accurate distance estimates (trgb, sbf) are marked with blue circles, objects with low precision 
distance estimates (NAM, TF, mem) are marked with black crosses. The new dwarf galaxies detected by us are highlighted in red. 
 
 All six of our galaxies belong to the BCD and dIrr types with active star formation, which is characteristic of galaxies located in a 
poor environment. Together with the BCD galaxy NGC\,6789, they form a more or less isolated diffuse association of $\sim2$~Mpc size. 
There may be other faint dwarf galaxies in this region at lower Galactic latitudes that are not covered by the Legacy survey area. 
Examples of other dwarf galaxy associations in the Local Volume have been presented by \citet{Tul2006}.
 
 \section{Concluding remark} Thanks to recent massive sky surveys in the optical range (DESI Legacy Imaging Surveys) and in the HI 21~cm 
line (ALFALFA, FASHI), many nearby dwarf galaxies with active (dIrr, BCD) and quiescent (dSph) star formation have been detected. Most of 
them do not have reliable distance estimates. Radial velocities have been measured for a very small number of dSph galaxies, as these 
measurements require a significant investment of observational time. Forthcoming surveys of the sky beyond the Earth's atmosphere promise to 
detect hundreds more faint dwarf galaxies in the Local Volume. Determining their distances and radial velocities will provide a unique basis 
for measuring the masses of local attractors on small cosmological scales, which is relevant for improving the standard $\Lambda$CDM model.

 \section{DATA AVAILABILITY} The data on which this work is based is publicly available and is detailed in the corresponding tables of 
the manuscript.

 \section{Acknowledgements}
 The authors thank the anonymous referee for the useful comments.
We are grateful to R.I.Uklein for assistance in observations and to A.V.Moiseev for useful advices. This work has made use of DESI Legacy 
Imaging Surveys data, and the revised version of the Local Volume galaxies database. The LV database has been updated within the framework 
of grant of the Russian Science Foundation 
No 24-12-00277. 
 
  %\bibliographystyle{mnras}
%\bibliography{Karachentsev_5}
%\end{document}

 \begin{table*} 
 \caption{Radial velocities of 40 nearby dwarf galaxies observed with the 6-m BTA telescope
.}
\begin{tabular}{lclrrlr} \hline
   Name       &     RA (2000.0) Dec.   &    $B$   &      $V_h$    &   $V_{\rm LG}$  & lines & $D_{\rm NAM}$ \\ 
\hline                                                                         
 (1)          &    (2)                &  (3)   &      (4)   &   (5)   &  (6)  & (7)  \\
 \hline                                                                        
 dw0136+1628  &  01 36 20.2 +16 28 12 &   18.2 &    550$\pm$60 &    721  &   abs &  9.1
 \\
 NGC628dwA    &  01 37 17.8 +15 37 58 &   18.6 &    497$\pm$50 &    666  &   abs &  8.5
 \\
 N672dw0149+27&  01 49 49.7 +27 30 50 &   18.0 &   3389$\pm$32 &   3585  &   abs & 38.6 
\\
 KKH 11       &  02 24 35.0 +56 00 42 &   16.2 &    111$\pm$10 &    343  &   em  &  6.1
 \\
 Dw0742+77    &  07 42 07.7 +77 41 28 &   17.5 &   1406$\pm$10 &   1593  &   em  & 23.0
 \\
 Dw0831+77    &  08 31 59.8 +77 36 07 &   17.  &   1428$\pm$10 &   1611  &   em  & 23.5
 \\
 PGC 025409   &  09 02 50.6 +71 18 22 &   15.8 &    416$\pm$10 &    569  &   em  & 10.8
 \\
 UGC 4918     &  09 19 17.7 +69 48 04 &   15.7 &    710$\pm$41 &    870  &   abs & 14.9 
\\
 Dw0921+75    &  09 21 36.0 +75 16 41 &   19.0 &    619$\pm$25 &    790  &   em  & 13.5 
\\
 SMDG0956+82  &  09 56 13.0 +82 53 24 &   18.1 &    $-$63$\pm$10 &  142  &   em  &  2.6
 \\
 MCG+13-07    &  09 56 26.4 +78 51 14 &   15.5 &   1292$\pm$10 &   1479  &   em  & 21.7
 \\
 Dw1012+42    &  10 12 42.7 +42 59 31 &   17.9 &   2326$\pm$47 &   2330  &   abs & 37.2
 \\
 KDG 74       &  11 02 21.8 +70 15 50 &   18.6 &    208$\pm$48 &    360  &   abs &  7.4
 \\
 Dw1119+10    &  11 19 30.7 +10 11 56 &   18.1 &   1029$\pm$10 &    879  &   em  & 10.3
 \\
 UGC 6451     &  11 28 46.4 +79 36 07 &   15.4 &     33$\pm$20 &    228  &   em  &  3.6
 \\
 $[$KK98$]$121    &  12 05 24.5 +43 42 28 &   15.2 &    689$\pm$51 &    725  &   abs & 10.3
 \\
 Dw1234+76    &  12 34 23.3 +76 43 34 &   18.0 &    202$\pm$76 &    393  &   abs &  7.2
 \\
 KDG 162      &  12 35 01.6 +58 23 08 &   17.3 &    105$\pm$50 &    222  &   abs &  2.8
 \\
 Dw1238+33    &  12 38 18.0 +33 37 59 &   17.7 &   1083$\pm$10 &   1085  &   em  & 18.7
 \\
 SMDG1241+35  &  12 41 11.0 +35 11 46 &   18.7 &    661$\pm$10 &    672  &   em  &  7.8
 \\
 Dw1245+61    &  12 45 49.0 +61 58 08 &   18.4 &     68$\pm$40 &    204  &   abs &  2.7
 \\
 Dw1339+39    &  13 39 45.1 +39 08 09 &   17.0 &    679$\pm$10 &    742  &   em  &  9.8
 \\
 Dw1352+61    &  13 52 39.4 +61 42 50 &   17.4 &   2105$\pm$10 &   2264  &   em  & 34.2
 \\
 Dw1536+81    &  15 36 06.5 +81 23 42 &   17.  &   1502$\pm$10 &   1731  &   em  & 24.7
 \\
 Dw1538+59    &  15 38 08.2 +58 59 56 &   17.9 &   2937$\pm$10 &   3135  &   em  & 45.1
 \\
 Dw1558+67    &  15 58 46.8 +67 51 29 &   16.8 &    $-$25$\pm$10 &  207  &   em  &  3.1
 \\
 Dw1559+46    &  15 59 02.6 +46 23 40 &   17.1 &     67$\pm$10 &    247  &   em  &  3.4
 \\
 Dw1645+46    &  16 45 48.5 +46 47 24 &   17.6 &    $-$21$\pm$10 &  183  &   em  &  2.9
 \\
 VV 507       &  17 00 42.2 +53 21 36 &   16.0 &   1122$\pm$10 &   1352  &   em  & 20.6
 \\
 Dw1707+51    &  17 07 31.7 +51 56 53 &   18.  &   1029$\pm$20 &   1261  &   em  & 19.3
 \\
 Dw1709+74    &  17 09 45.6 +74 10 44 &   17.2 &   1298$\pm$10 &   1544  &   em  & 22.0  \\
 Dw1717+48    &  17 17 19.9 +48 03 04 &   16.0 &   2421$\pm$10 &   2653  &   em  & 36.9
 \\
 Dw1726+80    &  17 26 30.7 +80 43 23 &   17.5 &   1811$\pm$10 &   2056  &   em  & 28.8  \\   
 Dw1735+57    &  17 35 34.6 +57 48 47 &   16.8 &     42$\pm$10 &    293  &   em  &  4.6
 \\
 Dw1826+44    &  18 26 39.1 +44 07 26 &   18.  &   2353$\pm$10 &   2617  &   em  & 34.1
 \\
 Dw1902+63    &  19 02 27.8 +63 48 40 &   17.  &   2950$\pm$10 &   3232  &   em  & 40.3 
\\
 Dw1905+65    &  19 05 42.7 +65 55 37 &   18.  &    712$\pm$10 &    992  &   em  & 14.6
 \\
 Dw1907+63    &  19 07 15.4 +63 23 06 &   17.  &   $-$147$\pm$10 &  136  &   em  &  2.4
 \\
 Dw1940+64    &  19 40 55.4 +64 45 32 &   18.  &    $-$59$\pm$10 &  231  &   em  &  3.8 \\  
 EZOAJ2120+45 &  21 20 46.2 +45 16 22 &   16.3 &    269$\pm$10 &    584  &   em  &  7.9 \\
 \hline
 \end{tabular}
 \end{table*}

 \begin{table*}
  \caption{LV galaxies with $D < 7.0$ Mpc and   SGZ $> 1.5$ Mpc.}
  \begin{tabular}{lrrccllrrcc}
 \hline                           
     Name         &         RA  &  Dec   &    SGL  &     SGB &     $D$  &  meth &  SGX  &   SGY  &    SGZ  & Ref.\\
\hline                                                                                          
     (1)          &       (2)   &   (3)  &  (4)    &   (5)   &    (6) &   (7) & (8)   &  (9)   &   (10) & (11) \\    \hline
 AGC748778        &      1.643  & 15.510 & 308.592 & 15.486  &   6.22 &  TRGB & 3.739 & $-$4.685 &  1.661  & [1]\\
UGC00288         &      7.266  & 43.431 & 338.004 & 15.374  &   6.82 &  TRGB & 6.097 & $-$2.463 &  1.808  & [1] \\ 
PGC2229803       &    201.971  & 43.815 &  74.649 & 16.662  &   6.58 &  NAM  & 1.669 &  6.079 &  1.887   &  --- \\
KK207            &    203.357  & 56.500 &  61.429 & 18.325  &   6.95 &  mem  & 3.155 &  5.794 &  2.185   &  ---\\
dw1343+58        &    205.779  & 58.228 &  59.593 & 19.620  &   5.59 & TRGB  & 2.660 &  4.528 &  1.875   &  [1] \\  
 Dw1348+6004      &    207.023  & 60.066 &  57.609 & 20.175  &   6.95 &  mem  & 3.495 &  5.509 &  2.397   &  ---\\
dw1350+5441      &    207.742  & 54.688 &  63.324 & 20.820  &   6.27 &  sbf  & 2.631 &  5.237 &  2.229   &  [2]\\
Dw1351+5014      &    207.995  & 50.248 &  68.074 & 21.075  &   6.95 &  mem  & 2.421 &  6.016 &  2.499  &  ---\\
GBT 1355+5439    &    208.710  & 54.647 &  63.343 & 21.381  &   6.95 &  mem  & 2.903 &  5.784 &  2.534  &  ---\\
dw1355+51        &    208.795  & 51.908 &  66.282 & 21.543  &   6.95 &  mem  & 2.600 &  5.918 &  2.552  &  ---\\
UGC08882         &    209.310  & 54.100 &  63.910 & 21.759  &   6.95 &  sbf  & 2.839 &  5.797 &  2.576  &  [2]\\
Dw1358+5255      &    209.530  & 52.918 &  65.176 & 21.950  &   6.95 &  mem  & 2.706 &  5.851 &  2.598  &  ---\\
M101-df3         &    210.773  & 53.615 &  64.375 & 22.650  &   6.52 & TRGB  & 2.602 &  5.425 &  2.511  &  [3]\\
MESSIER101       &    210.803  & 54.350 &  63.579 & 22.613  &   6.95 & TRGB  & 2.855 &  5.746 &  2.672  &  [1]\\ 
M101-df1         &    210.937  & 53.944 &  64.011 & 22.722  &   6.37 & TRGB  & 2.575 &  5.281 &  2.461  &  [3]\\
NGC5474          &    211.258  & 53.663 &  64.300 & 22.933  &   6.98 & TRGB  & 2.788 &  5.792 &  2.720  &  [1]\\
NGC5477          &    211.387  & 54.460 &  63.430 & 22.943  &   6.76 & TRGB  & 2.784 &  5.568 &  2.635  &  [1]\\
M101-dwA         &    211.709  & 53.742 &  64.190 & 23.193  &   6.65 & TRGB  & 2.661 &  5.503 &  2.619  &  [4]\\
M101Dw7          &    211.837  & 55.064 &  62.751 & 23.145  &   6.95 &  mem  & 2.926 &  5.681 &  2.732  &   ---\\
M101-df2         &    212.156  & 54.325 &  63.532 & 23.401  &   6.87 & TRGB  & 2.810 &  5.644 &  2.729  &  [3]\\
Dw1409+5113      &    212.304  & 51.225 &  66.894 & 23.757  &   6.95 &  mem  & 2.496 &  5.851 &  2.800  &  ---\\
dw1416+57        &    214.245  & 57.910 &  59.486 & 24.105  &   6.95 &  mem  & 3.221 &  5.465 &  2.838  &  ---\\
NGC5585          &    214.951  & 56.730 &  60.701 & 24.665  &   7.00 & TRGB  & 3.108 &  5.550 &  2.919  &  [1]\\
DDO194           &    218.852  & 57.256 &  59.683 & 26.653  &   5.81 & TRGB  & 2.621 &  4.483 &  2.606  &  [1]\\
dw1446+58        &    221.750  & 58.567 &  57.825 & 27.818  &   6.95 &  mem  & 3.273 &  5.203 &  3.243  &  ---\\
NGC6503          &    267.365  & 70.144 &  33.138 & 34.633  &   6.25 & TRGB  & 4.306 &  2.811 &  3.552  &  ---\\
KK242            &    268.201  & 70.137 &  32.814 & 34.732  &   6.46 & TRGB  & 4.462 &  2.877 &  3.681  &  [5]\\
UGC11411         &    287.176  & 70.283 &  25.035 & 35.384  &   6.58 & TRGB  & 4.861 &  2.270 &  3.810  &  [1]\\
HIZOAJ1914+10    &    288.760  & 10.328 & 249.238 & 82.739  &   6.90 &  bTF  &$-$0.309 & $-$0.815 &  6.845  &  [6]\\
NGC6789          &    289.174  & 63.972 &  23.273 & 41.591  &   3.55 & TRGB  & 2.439 &  1.049 &  2.357  &  [1]\\
KK246            &    300.989  &-31.681 & 225.611 & 39.807  &   6.85 & TRGB  &$-$3.681 & $-$3.760 &  4.385  &  [1]\\
EZOAJ2129+52     &    322.267  & 52.654 & 355.774 & 42.235  &   6.35 &  NAM  & 4.689 & $-$0.346 &  4.268   &  ---  \\ 
 \hline
 Dw1558+67        &    239.695  & 67.858 &  44.210 & 30.764  &   3.1  &  NAM  & 1.909 &  1.858 &  1.586  &  ---\\
 Dw1559+46        &    239.760  & 46.394 &  66.858 & 42.352  &   3.4  &  NAM  & 0.987 &  2.311 &  2.291  &  ---\\
 Dw1645+46        &    251.452  & 46.790 &  60.328 & 48.976  &   2.9  &  NAM  & 0.943 &  1.654 &  2.187  &  ---\\
 Dw1735+57        &    263.894  & 57.813 &  41.372 & 45.351  &   4.6  &  NAM  & 2.425 &  2.136 &  3.272  &  ---\\
 Dw1907+63        &    286.814  & 63.385 &  24.598 & 42.276  &   2.4  &  NAM  & 1.615 &  0.644 &  1.614  &  ---\\
 Dw1940+64        &    295.230  & 64.758 &  20.060 & 40.330  &   3.8  &  NAM  & 2.721 &  0.993 &  2.459  &  ---\\
\hline
 \multicolumn{11}{l}{{\bf Notes}: [1]  \citet{Ana2021}, [2] \citet{Car2022}, [3] \citet{Dan2017},}\\
 \multicolumn{11}{l}{[4]  \citet{Kar2019}, [5] \citet{Kar2022b},  [6] \citet{Kar2013}.} \\
   \end{tabular}
   \end{table*}

   \end{document}